\documentclass[10pt,twocolumn,showpacs,amsmath,amssymb,floatfix,superscriptaddress]{revtex4-2}
\usepackage{graphicx}
\usepackage{float}
\usepackage{dcolumn}
\usepackage{bm}
\usepackage{color}
\usepackage{txfonts}
\usepackage{microtype}
\usepackage{hyperref}
\usepackage{mathrsfs}
\usepackage{easyReview}
\usepackage{amsmath}
\usepackage{extarrows}
\usepackage{mathtools}
\usepackage{xcolor}
\usepackage{slashed}
\usepackage{amsmath}
\usepackage{ulem}

\setreviewson
\begin{document}

\title{Generation of ultrarelativistic vortex leptons with large orbital angular momenta}

\author{Mamutjan Ababekri}
\affiliation{Ministry of Education Key Laboratory for Nonequilibrium Synthesis and Modulation of Condensed Matter, Shaanxi Province Key Laboratory of Quantum Information and Quantum Optoelectronic Devices, School of Physics, Xi'an Jiaotong University, Xi'an 710049, China}
\author{Jun-Lin Zhou}
\affiliation{Ministry of Education Key Laboratory for Nonequilibrium Synthesis and Modulation of Condensed Matter, Shaanxi Province Key Laboratory of Quantum Information and Quantum Optoelectronic Devices, School of Physics, Xi'an Jiaotong University, Xi'an 710049, China}
\author{Ren-Tong Guo}
\affiliation{Ministry of Education Key Laboratory for Nonequilibrium Synthesis and Modulation of Condensed Matter, Shaanxi Province Key Laboratory of Quantum Information and Quantum Optoelectronic Devices, School of Physics, Xi'an Jiaotong University, Xi'an 710049, China}
\author{Yong-Zheng Ren}
\affiliation{Ministry of Education Key Laboratory for Nonequilibrium Synthesis and Modulation of Condensed Matter, Shaanxi Province Key Laboratory of Quantum Information and Quantum Optoelectronic Devices, School of Physics, Xi'an Jiaotong University, Xi'an 710049, China}
\author{Yu-Han Kou}
\affiliation{Ministry of Education Key Laboratory for Nonequilibrium Synthesis and Modulation of Condensed Matter, Shaanxi Province Key Laboratory of Quantum Information and Quantum Optoelectronic Devices, School of Physics, Xi'an Jiaotong University, Xi'an 710049, China}
\author{Qian Zhao}
\affiliation{Ministry of Education Key Laboratory for Nonequilibrium Synthesis and Modulation of Condensed Matter, Shaanxi Province Key Laboratory of Quantum Information and Quantum Optoelectronic Devices, School of Physics, Xi'an Jiaotong University, Xi'an 710049, China}
\author{Zhong-Peng Li}
\affiliation{Ministry of Education Key Laboratory for Nonequilibrium Synthesis and Modulation of Condensed Matter, Shaanxi Province Key Laboratory of Quantum Information and Quantum Optoelectronic Devices, School of Physics, Xi'an Jiaotong University, Xi'an 710049, China}
\author{Jian-Xing Li}\email{jianxing@xjtu.edu.cn}
\affiliation{Ministry of Education Key Laboratory for Nonequilibrium Synthesis and Modulation of Condensed Matter, Shaanxi Province Key Laboratory of Quantum Information and Quantum Optoelectronic Devices, School of Physics, Xi'an Jiaotong University, Xi'an 710049, China}
\affiliation{Department of Nuclear Physics, China Institute of Atomic Energy, P. O. Box 275(7), Beijing 102413, China}
\date{\today}

\begin{abstract}
We put forward a novel method for generating ultrarelativistic vortex positrons and electrons with intrinsic orbital angular momenta (OAM) through nonlinear Breit-Wheeler (NBW) scattering of vortex $\gamma$ photons. A complete angular momentum-resolved scattering theory has been formulated, introducing  angular momenta of laser photons and vortex particles into the conventional NBW scattering framework. We find that vortex positron (electron) can be produced when the outgoing electron (positron) is generated along the collision axis. By unveiling the angular momentum transfer mechanism, we clarify that the OAM of the $\gamma$ photon and the large angular momenta due to multiple laser photons are entirely transferred to the generated vortex leptons. Furthermore, the cone opening angle and superposition state of the vortex $\gamma$ photon can be determined via the angular distribution of created pairs in NBW processes.  
\end{abstract}
\maketitle

\section{introduction}

Vortex particles with intrinsic orbital angular momenta (OAM) exhibit wave packets featuring helical phase fronts \cite{allen1992orbital,bliokh2017theory,lloyd2017electron,knyazev2018beams}.  They can introduce novelty to scattering processes owing to their distinctive properties, including large angular momenta (AM) and helical wave fronts \cite{ivanov2020doing,budker2021arxiv,ivanov2022promises,Lu:2023wrf}. Collision of ultrarelativistic vortex particle beams presents an opportunity to expand the scope of future colliders \cite{Benedikt:2020ejr,Shiltsev:2019rfl,Anderle:2021wcy}.  Utilizing ultrarelativistic vortex leptons in deep inelastic scattering experiments has the potential to provide new insight into the spin and OAM constituents of protons, offering potential solutions to the long-standing proton spin puzzle \cite{Aidala:2012mv,Leader:2013jra,Larocque2018,Vanacore2019}. Currently, vortex electrons can be produced with kinetic energy up to about 300 keV using spiral phase plates, fork diffraction gratings, or magnetic needles \cite{verbeeck2010production,Beche:2013wua,Grillo:2014ksg}.  Neutrons and atoms can also be brought into vortex states using similar wave front engineering techniques at low energies \cite{Clark:2015rcq,Alon:2021,Sarenac_2018,doi:10.1073/pnas.1906861116,sarenac2022experimental}. Unfortunately, the generation of ultrarelativistic ones remains a great challenge.  

In theory, charged particles can be initially prepared in vortex states at low energies and then accelerated to high energies \cite{Silenko:2017fvf,Silenko:2018eed,Silenko:2019ziz}. Their precise behavior during high-energy acceleration and the preservation of vortex structures in practical external fields require theoretical investigation and await experimental verification \cite{ivanov2022promises}. Another approach to producing high-energy vortex particles involves collision events, utilizing the entanglement of final state particles in scattering processes \cite{jentschura2011generation,jentschura2011compton,ivanov2012creation,van2015inelastic,Karlovets:2022evc,Karlovets:2022mhb}. In this scenario, one final particle can acquire a vortex state if specific postselection procedures are applied to the other. For instance, a generalized measurement can be employed and measures the azimuth angle of the final particle with significant uncertainty to induce the vortex state in the other particle \cite{Karlovets:2022evc}. However, generalized measurement is known only for low-energy processes \cite{10.1093/oso/9780198527626.001.0001,2022OExpr..3022396A}, and implementing it in high-energy collisions poses substantial challenges. Another possibility is demonstrated for the Compton scattering process, in which the helical wave front is transferred from the initial vortex photon to the final vortex $\gamma$ photon if the final electron is scattered along the collision axis \cite{jentschura2011generation}.  Lepton-antilepton pairs can be created through the conversion of light into matter, as demonstrated in the Breit-Wheeler (BW) or Bethe-Heitler (BH) scattering processes \cite{Breit:1934zz,Bethe:1934za,Motz:1969ti,Tsai:1973py,Klein:2004is,ATLAS:2017fur,STAR:2019wlg,Brandenburg:2022tna}. There are proposals for generating vortex pairs via linear BW and BH processes by assuming a twisted detector that performs coherent vortex postselection over final particles \cite{Bu:2021ebc,Lei:2021eqe,Lei:2023wui}. However, the possibility of generating vortex leptons through a feasible postselection scenario with a considerable probability in pair creation dynamics remains unknown. \\

Meanwhile, using ultrashort and ultraintense laser pulses  \cite{Weber:2017pmh,doi:10.1126/science.359.6374.382,danson2019petawatt,yoon2019achieving,yoon2021realization}, the nonlinear Breit-Wheeler (NBW) scattering has emerged as a promising method for generating high-brilliance and high-energy positron beams \cite{DiPiazza:2011tq,Gonoskov:2021hwf,Fedotov:2022ely,Bulanov:2023cxq}.  Recent advancements in this field have also revealed the related polarization features, allowing for the production of polarized pairs \cite{Vranic:2018liw,Chen:2019vly,Wan:2019gow,Li:2020bwo,Büscher_Hützen_Ji_Lehrach_2020,Song:2021tkc,Dai:2021vgl,Sun:2022cat,Xue:2023xwg}. However, it is worth noting that current studies focus on investigating the spin angular momentum (SAM) as the only internal degree of freedom \cite{Ivanov:2004vh,RevModPhys.87.247,Jansen:2016gvt,Seipt:2020diz,Torgrimsson:2020gws,Chen:2022dgo,Tang:2022tmn,Zhuang:2023lwo}. Furthermore, since NBW pair creation entails the absorption of multiple laser photons, a substantial amount of AM carried by photons in the laser field is expected to be transferred to the generated pairs. While the overall beam in NBW processes may acquire extrinsic (mechanical) OAM \cite{Chen:2018tkb,Zhu:2018und,Zhao2022,Zhang:2024ofv}, the intrinsic OAM carried by scattering particles ($\gamma$ photons, electrons, and positrons) remain unknown. 
Consequently, the AM transfer mechanism and the formation of a helical wave front in the final state during NBW processes are still unclear, leaving the generation of ultrarelativistic vortex leptons in intense lasers an open question.

\begin{figure}[H]
	\setlength{\abovecaptionskip}{-0.0cm}
	\includegraphics[width=0.975\linewidth]{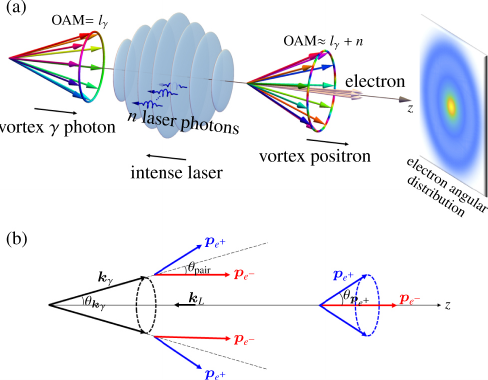}
	\begin{picture}(300,0)	
	\end{picture}
	\caption{(a) Generation of an ultrarelativistic vortex positron with large OAM via the NBW process. A vortex $\gamma$ photon carrying OAM $l_\gamma$ collides head-on with a CP laser and absorbs $n$ laser photons, subsequently decaying into electron and positron pair. When the electron is postselected along the collision axis, the generated positron assumes a vortex state with OAM $l_\gamma+n$. The angular distribution exhibits a bright spot, indicating a maximum probability for electron creation along the collision axis.  (b) The condition for producing a vortex positron illustrated in momentum space. The momentum vectors $\bm{k}_\gamma$, $\bm{k}_L$, $\bm{p}_{e^-}$, and $\bm{p}_{e^+}$ correspond to the $\gamma$ photon, laser photon, electron, and positron, respectively. $\theta_{\bm{k}_\gamma}$ is the cone opening angle of the vortex state, $\theta_{\bm{p}_{e^+}}$ is the polar angle of the momentum vector $\bm{p}_{e^+}$, and $\theta_{\text{pair}}$ is the NBW pair creation angle relative to the $\gamma$ photon wave vector $\bm{k}_\gamma$.} 
	\label{fig_1}
\end{figure}

In this work, we investigate the generation of ultrarelativistic vortex positrons and electrons via NBW scattering of a vortex $\gamma$ photon in a circularly polarized (CP) laser [Fig.~\ref{fig_1} (a)]. We develop the AM-resolved NBW scattering theory within the strong-field quantum electrodynamics (SF-QED) framework \cite{Ritus:1985vta,DiPiazza:2011tq,Fedotov:2022ely} while incorporating the AM of all participating particles: laser photons, vortex $\gamma$ photon, and the generated leptons. We establish the vortex positron generation condition by examining the kinematics in momentum space, specifically by exploring the possibility of postselecting a plane-wave electron to generate a vortex positron. We find that vortex positrons arise upon postselecting electrons along the collision axis ($\bm{p}_{e^-} \parallel \hat{\bm{z}}$), and this scenario can be optimized to yield a considerable probability. Under the condition $\bm{p}_{e^-} \parallel \hat{\bm{z}}$, we derive the AM conservation relation and analyze impacts of multiphoton absorption and total SAM from scattering particles. We find that the OAM of the positron receives contributions from the $\gamma$ photon OAM and is amplified by the AM of multiple laser photons. These findings also hold for the symmetric case: ultrarelativistic vortex electrons can be generated when the positron is created along the collision axis.\\

Our paper is organized as follows: In Sec. \uppercase\expandafter{\romannumeral2}, we introduce basic considerations of NBW scattering of the vortex \(\gamma\) photon and establish the condition for the generation of vortex pairs. In Sec. \uppercase\expandafter{\romannumeral3}, we analyze the properties of the vortex pairs by presenting the theoretical framework for vortex positron generation and obtaining numerical results for the pair creation rates. We discuss the feasibility of our proposal and the detection of the vortex $\gamma$ photon in Sec.\uppercase\expandafter{\romannumeral4}. Finally, we end our paper with a brief summary in Sec. \uppercase\expandafter{\romannumeral5}.\\

Throughout, natural units are used ($\hbar=c=1$), the fine structure constant is \(\alpha=\frac{e^2}{4\pi}\approx\frac{1}{137}\) with the electron charge \(e=-\vert e\vert\), and the electron mass is denoted as \(m_e\). \\

\section{NBW scattering of the vortex $\gamma$ photon}

We consider the head-on collision between a vortex \(\gamma\) photon and a plane-wave laser pulse along the \(z\) axis [Fig.~\ref{fig_1} (a)] and investigate the pair creation process by calculating the respective NBW scattering within the Furry picture of SF QED \cite{Ritus:1985vta,DiPiazza:2011tq,Fedotov:2022ely}. In the following, we first introduce our treatment of the incoming vortex \(\gamma\) photon and then establish the kinematic condition for obtaining final state vortex pairs.\\ 

\subsection{Kinematic condition for the generation of vortex pairs} 

The vortex $\gamma$ photon propagating along the $z$ axis can be described by the Bessel wave packet in terms of the plane-wave function $A^\mu_{\Lambda}(t,\bm{r})$ as \cite{jentschura2011generation}
\begin{equation}
\mathcal{A}^{\mu}_{j_\gamma\,\Lambda}(t,\bm{r})=\int\frac{d^2\bm{k}_{\gamma,\perp}}{(2\pi)^2}\,a_{j_\gamma}(\bm{k}_{\gamma,\perp})\,A^\mu_{\Lambda}(t,\bm{r})\,,
\end{equation}   
where $\Lambda$ represents the helicity, $j_\gamma$ is the total angular momentum (TAM), and the vortex amplitude $a_{j_\gamma}(\bm{k}_{\gamma,\perp})$  reads
\begin{equation}
	a_{j}(\bm{k}_{\gamma,\perp})=(-i)^{j}\,e^{i\,j\,\varphi_{\bm{k}_\gamma}}\,\sqrt{\frac{2\pi}{k_{\gamma,\perp0}}}\delta(\vert\bm{k}_{\gamma,\perp}\vert-k_{\gamma,\perp0})\,.
	\label{eq_a}
\end{equation} 
The vortex amplitude encodes a cone structure with radius $\vert \bm{k}_{\gamma,\perp} \vert =k_{\gamma,\perp0}$ in the momentum space featuring a helical phase factor $e^{i\,j_\gamma\varphi_{\bm{k}_\gamma}}$. In the illustration shown in Fig.~\ref{fig_1}, we represent a Bessel vortex particle using a cone with an opening angle in momentum space. In general, the Bessel mode wave function of the vortex particle $\vert\psi_\text{vortex}\rangle$ is expressed in terms of the plane-wave function $\vert\psi_\text{plane}\rangle$ by employing the vortex amplitude in Eq.~\eqref{eq_a} as $\vert\psi_\text{vortex}\rangle=\int\,\frac{d^2\bm{p}_\perp}{(2\pi)^2}\,a_{j}(\bm{p}_\perp)\vert\psi_\text{plane}\rangle$. Bessel modes represent the simplest kink of vortex modes and allow for detailed calculations while reflecting key physics aspects related to AM transfer mechanism properly. The generation of vortex pairs involves obtaining similar wave packet structures for final states during NBW scattering.  \\

The condition for generating the vortex positron and electron is determined through the following argument in momentum space [Fig.~\ref{fig_1} (b)]. The momentum vectors $\bm{k}_\gamma$ of the constituent plane waves in a vortex photon exhibit a conical distribution characterized by an opening angle $\theta_{\bm{k}_\gamma}$ given by $\theta_{\bm{k}_\gamma} = \arcsin(\lvert\bm{k}_{\gamma,\perp}\rvert/\lvert\bm{k}_\gamma\rvert)$, where $\bm{k}_{\gamma,\perp}$ is the transverse component of the momentum $\bm{k}_\gamma$ \cite{jentschura2011generation}. A generated positron can be identified as a Bessel mode vortex particle when it exhibits a similar conical structure in momentum space \cite{Bliokh:2011fi,Hayrapetyan:2014faa}. For each momentum vector $\bm{k}_\gamma$, the generated electron-positron pair possesses an angle $\theta_{\text{pair}}$ with respect to the propagation direction of the plane-wave component.  Under the condition $\theta_{\text{pair}}=\theta_{\bm{k}_\gamma}$, the plane-wave electron is produced along the collision axis, with a zero polar angle with respect to this axis ($\theta_{\bm{p}_{e^-}}=0$), and the momentum vector $\bm{p}_{e^+}$ of the plane-wave positron is uniquely determined, exhibiting a fixed polar angle $\theta_{\bm{p}_{e^+}}$. As the wave vector $\bm{k}_\gamma$ traverses the cone, the momentum $\bm{p}_{e^+}$ of the positron also spreads out over a cone with an opening angle $\theta_{\bm{p}_{e^+}}$, thereby indicating its vortex state.  Note that the polar angle $\theta_{\bm{p}_{e^-}}$  could take a very small, nonzero value, and the vortex positron can be generated as long as the electron's azimuthal angle $\varphi_{\bm{p}_{e^-}}$ remains undetermined (see Appendixes \ref{Appendix_A} and \ref{Appendix_B}). These arguments also apply to the symmetric case. When the positron is created along collision axis ($\theta_{\bm{p}_{e^+}}=0$), the electron momentum $\bm{p}_{e^-}$ lies on a cone, resulting in the generation of a vortex electron.  In the following, we continue  our exploration of vortex positron generation. 

\subsection{Angular distribution of the created pairs}

As postselection of an electron along the collision axis is critical for vortex positron generation, we examine the angle-resolved probability distribution of the created pairs. The NBW scattering of the vortex $\gamma$ photon takes place in a CP plane-wave field,
\begin{eqnarray}
	\begin{aligned}
		A^\mu(\phi=k_L\cdot x)=a\,g(\phi)
		\begin{pmatrix} 0 \\ \cos\phi \\ \sin\phi \\ 0 \end{pmatrix}\,,
	\end{aligned}\label{laser_field}
\end{eqnarray}
where $k_L$ denotes the momentum four-vector of the laser photon, $a=a_0{m_e}/{e}$ is the normalized vector potential with the dimensionless parameter $a_0$, and $g(\phi)$ denotes the pulse shape function. The $S$-matrix element for the creation of plane-wave pairs from the vortex $\gamma$ photon during NBW processes ({\rm $\gamma_{\text{vortex}}$ + $n\,\omega_L$ $\rightarrow$ $e^+_{\text{plane wave}}$ + $e^-_{\text{plane wave}}$}) can be expressed as
\begin{equation}
	S_{fi}^\text{1vortex}=\int\frac{d^2\bm{k}_{\gamma,\perp}}{(2\pi)^2}a_{j_\gamma}(\bm{k}_{\gamma,\perp})\,S_{fi}^\text{plane}\,,
\end{equation}
where $S^{\text{plane}}_{fi}$ is the usual NBW $S$-matrix element with plane-wave particles \cite{Fedotov:2022ely}.  
The corresponding pair creation probability in terms of the electron's energy $\varepsilon_{e^-}$ and solid angle $\Omega_{e^-}$ is
\begin{equation}
	\frac{d^{2}W^{\text{1vortex}}}{d\varepsilon_{e^-}d\Omega_{e^-}}=\int\frac{d\varphi_{\bm{k}_\gamma}}{2\pi}\frac{d^2W^{\text{plane}}(\theta_{\bm{k}_\gamma},\varphi_{\bm{k}_\gamma})}{d\varepsilon_{e^-}d\Omega_{e^-}},
	\label{eq_vortex_pair_rate}
\end{equation}
where $\frac{d^2W^{\text{plane}}(\theta_{\bm{k}_\gamma},\varphi_{_{\bm{k}_\gamma}})}{d\varepsilon_{e^-}d\Omega_{e^-}}$ is the usual NBW pair creation probability for plane-wave particles in pulsed lasers \cite{Baier:1998vh,Wistisen:2020rsq}. 
The angle-differential probability ${dW^{\text{1vortex}}}/{d\Omega_{e^-}}$ is obtained by integrating out the energy. \\

\begin{figure}[H]
	\setlength{\abovecaptionskip}{-0.0cm}
	\includegraphics[width=0.95\linewidth]{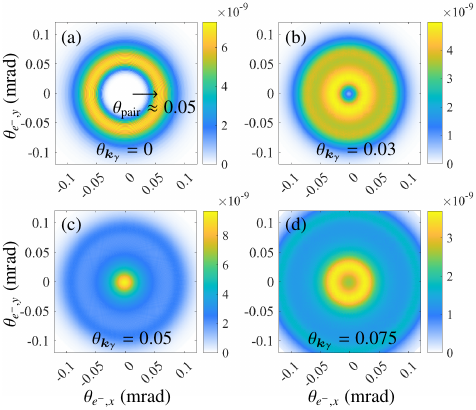}
	\begin{picture}(300,0)	 
	\end{picture}
	\caption{The probability distribution ${dW^{\text{1vortex}}}/{d\Omega_{e^-}}$ of the electron in $\theta_{e^-,x}\theta_{e^-,y}$ plane corresponding to NBW scattering in CP lasers with (a) the plane-wave $\gamma$ photon and (b)-(d) the vortex state $\gamma$ photons with different cone opening angles $\theta_{\bm{k}_\gamma}$. $\theta_{\bm{k}_\gamma}$ is the vortex $\gamma$ photon's cone opening angle. $\theta_{\bm{k}_\gamma}=0$ corresponds to the case of plane-wave $\gamma$ photon scattering with the laser along the collision axis.} 
	\label{fig_CP}
\end{figure} 

\begin{figure}[H]
	\setlength{\abovecaptionskip}{-0.0cm}
	\includegraphics[width=0.9\linewidth]{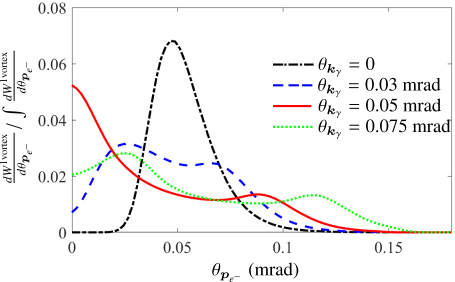}
	\begin{picture}(300,0)	
	\end{picture}
	\caption{Normalized probability distribution vs electron's propagation angle $\theta_{\bm{p}_{e^-}}$ with respect to the collision axis. The distribution is normalized by the total probability $\int \frac{dW^{\text{1vortex}}}{d\theta_{\bm{p}_{e^-}}}$ and presented with the resolution $\Delta \theta_{\bm{p}_{e^-}}\approx0.002$ mrad.  } 
	\label{fig_02}
\end{figure}

In Fig.~\ref{fig_CP}, we present the probability distribution ${dW^{\text{1vortex}}}/{d\Omega_{e^-}}$ of the created electron in the $\theta_{e^-,x}\theta_{e^-,y}$ plane, obtained by using Eq.~\eqref{eq_vortex_pair_rate}, where we define $d\Omega_{e^-}=d\theta_{e^-,x}d\theta_{e^-,y}$. We set laser intensity to $a_0=1$ and employ the pulse shape function $g(\phi)=\sin^2(\frac{\phi}{2 N_{\text{cycle}}})$, where the number of laser pulse cycles is set as $N_{\text{cycle}}=8$, and the central laser frequency is $\omega_L=1.55$ eV. The energy of the vortex $\gamma$ photon is set to $\varepsilon_\gamma=20$ GeV, and calculations are conducted for various values of $\theta_{\bm{k}_\gamma}$. When electron-positron pairs are created from plane-wave $\gamma$ photons ($\theta_{\bm{k}_\gamma}=0$), the angle distribution features a ring with radius $\theta_{\text{pair}}\approx 0.05$ mrad [Fig.~\ref{fig_CP} (a)]. Here, $\theta_{\text{pair}}\sim a_0 m_e/\varepsilon_\gamma$ is the NBW  pair creation angle with the maximum probability.  When the vortex $\gamma$ photon is considered, the results depend on the relation between $\theta_{\bm{k}_\gamma}$ and $\theta_{\text{pair}}$, as shown in Figs.~\ref{fig_CP} (b)-2(d). In particular, when $\theta_{\text{pair}}=\theta_{\bm{k}_\gamma}$ [Fig.~\ref{fig_CP} (c)], there is a peak at the center of the angle-resolved probability distribution, which favors the special condition $\theta_{\bm{p}_{e^-}}=0$ for the generation of the vortex positron.  \\

In Fig.~\ref{fig_02}, we display the redefined angle-resolved result ($dW/d\theta_{\bm{p}_{e^-}}$) from Fig.~\ref{fig_CP} for direct comparison of results obtained for different $\theta_{\bm{k}_\gamma}$, where we introduce $\theta_{\bm{p}_{e^-}} = \sqrt{\theta_{e^-,x}^2 + \theta_{e^-,y}^2}$. The electrons generated along the collision axis fall within the angular range $0 \leq \theta_{\bm{p}_{e^-}} \leq 0.002$ mrad ($0 \leq \theta_{\bm{p}_{e^-}} \leq 0.005$ mrad), constituting 5.2\% (11.6\%) of the total electron yield, as illustrated by the solid red line in Fig.~\ref{fig_02}.  
In the plane-wave limit of the $\gamma$ photon ($\theta_{\bm{k}_\gamma}=0$), the electron emerges predominantly at an angle of $\theta_{\bm{p}_{e^-}}\approx0.05$ mrad relative to the collision axis, as evidenced by the peak position of the dot-dashed black line in Fig.~\ref{fig_02}.  Therefore, optimizing the probability for electron generation along the collision axis occurs when the cone opening angle of the vortex $\gamma$ photon aligns with the dominant pair creation angle associated with the plane-wave component of the $\gamma$ photon.\\

\section{Properties of the generated vortex pairs}\label{sec_3}

We obtain the vortex positron generation probability and elucidate the AM transfer mechanism by applying the kinematic condition $\theta_{\bm{p}_{e^-}}=0$, which leads to the generation of vortex positrons and yields the maximum probability, as discussed in the previous section. To this end, we develop an AM-resolved NBW scattering theory and numerically investigate the energy and OAM distribution properties of the generated vortex positron.\\
	
\subsection{Theoretical considerations}

\subsubsection{The AM-resolved NBW scattering theory}

To fully understand the AM transfer mechanism in NBW scattering, it is essential to take into account both the SAM and OAM contributions from all the particles involved. 
The harmonic expansion of the scattering amplitude is carried out to account for the SAM contribution resulting from multiple laser photon absorption in a CP laser. In the pulsed laser case, the harmonic expansion is possible via the slowly varying envelope approximation (SVEA) \cite{seipt2016analytical} assuming that the laser pulse is long enough ($N_{\text{cycle}}\gg1$) or that the pulse shape function is slowly varying [$\frac{\partial}{\partial \phi} g(\phi) \ll g(\phi)$]. After applying SVEA relations such as
$\int d\phi g(\phi) \sin(\phi) \approx -g(\phi) \cos(\phi)$ and $\int d\phi g(\phi) \cos(\phi) \approx g(\phi) \sin(\phi)$, one can perform the harmonic expansion using the Jacobi-Anger type expansion
\begin{equation}
	e^{i\alpha g(\phi)\sin(\phi + \varphi_{\bm{p}})}=\sum_n J_n(\alpha g(\phi)) e^{i n ({\phi} + \varphi_{\bm{p}})}\,,
	\label{harmonic_expansion}
\end{equation}
where $J_n$ is the Bessel function of the first kind. In this context, $\phi$ represents the laser phase, whereas $\varphi_{\bm{p}}$  denotes the azimuthal angle of the electron/positron's momentum.\\

\begin{figure}[H]
	\setlength{\abovecaptionskip}{-0.0cm}
	\includegraphics[width=1\linewidth]{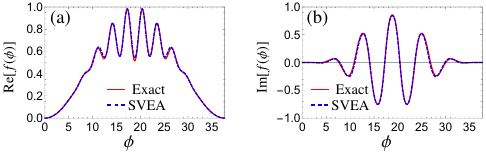}
	\begin{picture}(300,0)	
	\end{picture}
	\caption{Comparison of the exact result for $f(\phi)=g(\phi) e^{-i\int d\phi g(\phi) \sin \phi}$ (solid red curves) and the corresponding SVEA result (dashed blue curves), where real and imaginary parts are presented in (a) and (b), respectively. The pulse shape function is $g(\phi)=\sin^2(\frac{\phi}{2 N_{\text{cycle}}})$ with $N_{\text{cycle}}=6$.} 
	\label{sfig_1}
\end{figure}

Regarding the numerical calculations in our work, it is observed that the use of the SVEA provides a satisfactory approximation, even for $N_{\text{cycle}}=6$, as demonstrated in Fig.~\ref{sfig_1}. For the optical laser ($\omega_L=1.55$ eV), this requirement translates to the laser pulse duration as, $\tau\geq16$ fs. The SVEA applies more to increasing values of $N_{\text{cycle}}$.\\ 

The electron and positron spinors, denoted as $u_{s_{e^-}}$ and $v_{s_{e^+}}$ respectively, along with the polarization vector of the $\gamma$ photon, denoted as $\bm{\epsilon}_{\Lambda}$, are expressed as expansions using the corresponding spin eigenfunctions
\begin{eqnarray}
	\begin{aligned}
		&u_{s_{e^-}}=\sum_{\sigma_{e^-}=\pm\frac{1}{2}}e^{-i\sigma_{e^-}\varphi_{\bm{p}_{e^-}}}d_{\sigma_{e^-},s_{e^-}}^{1/2}(\theta_{\bm{p}_{e^-}})\begin{pmatrix} \sqrt{\varepsilon_{e^-}+m_e}\,w^{\sigma_{e^-}}_{z} \\ 2s_{e^-}\sqrt{\varepsilon_{e^-}-m_e}\,w^{\sigma_{e^-}}_{z}\\ \end{pmatrix}, \\
		&v_{s_{e^+}}=\sum_{\sigma_{e^+}=\pm\frac{1}{2}}e^{i\sigma_{e^+}\varphi_{\bm{p}_{e^+}}}d_{\sigma_{e^+},s_{e^+}}^{1/2}(\theta_{\bm{p}_{e^+}})\begin{pmatrix} -2s_{e^+}\sqrt{\varepsilon_{e^+}-m_e}\,\tilde{w}^{\sigma_{e^+}}_{z} \\ \sqrt{\varepsilon_{e^+}+m_e}\,\tilde{w}^{\sigma_{e^+}}_z    \\   \end{pmatrix},  \\		
		&\bm{\epsilon}_{\Lambda}=\sum_{\sigma_{\gamma}=0,\pm1}e^{-i\sigma_\gamma\varphi_{\bm{k}_\gamma}}d_{\sigma_{\gamma},\Lambda}^{1}(\theta_{\bm{k}_\gamma})\bm{\chi}_{\sigma_\gamma}\,,
	\end{aligned}\label{SAM_basis_expansion}
\end{eqnarray}
where $\varepsilon_{e^-}$ ($\varepsilon_{e^+}$) is the electron (positron) energy. The variables $\varphi_{\bm{p}_{e^-}}$, $\theta_{\bm{p}_{e^-}}$, $\varphi_{\bm{p}_{e^+}}$, $\theta_{\bm{p}_{e^+}}$, $\varphi_{{\bm{k}_\gamma}}$, and $\theta_{\bm{k}_\gamma}$ represent the azimuthal and polar angles of the electron ($\bm{p}_{e^-}$), positron ($\bm{p}_{e^+}$), and $\gamma$ photon ($\bm{k}_\gamma$) momentum vectors, respectively. The spin eigenfunctions in Eq.~\eqref{SAM_basis_expansion} satisfy $\hat{s}_{z} w^{\sigma_{e^-}}_z=\sigma_{e^-} w^{\sigma_{e^-}}_z$ and $\hat{s}_{z}\bm{\chi}_{\sigma_{\gamma}}=\sigma_{\gamma} \bm{\chi}_{\sigma_{\gamma}}$, with the corresponding eigenvalues $\sigma_{e^-}=\pm 1/2$ and $\sigma_{\gamma}=\pm 1, 0$, respectively. The corresponding SAM are denoted by $s_{e^-}$, $s_{e^+}$, and $\Lambda$. The spinors are given by $ w^{1/2}_{z}=(1,0)^T$ and $ w^{-1/2}_{z}=(0,1)^T$, and the vectors are given by $\bm{\chi}_0=(0,0,1)^T$ and $\bm{\chi}_{\pm}=\frac{\mp}{\sqrt{2}}(1,\pm i,0)^T$.  The corresponding spin operators in the $z$ direction, $\hat{s}_z$, are a $2\times2$ Pauli matrix $\hat{s}_z=\frac{1}{2}\hat{\sigma}_z$ for the spinor and a $3\times3$ spin-1 matrix for the vector \cite{Bliokh:2015doa}. The spinors for the electron and the positron satisfy the relation $\tilde{w}^\pm_z=\pm w^\mp_z$.   
The Wigner's $d$ functions are 
\begin{eqnarray}
	\begin{aligned}
		d_{\sigma,s}^{1/2}(\theta)=\begin{pmatrix}
			\cos\theta/2 \qquad -\sin\theta/2 \\ \sin\theta/2 \qquad \cos\theta/2 \,
		\end{pmatrix}\,,
	\end{aligned}
\end{eqnarray}
and
\begin{eqnarray}
	\begin{aligned}
		d_{\sigma,\Lambda}^{1}(\theta)=\begin{pmatrix}
			\cos^2\theta/2 \qquad -\frac{1}{\sqrt{2}}\sin\theta \qquad \sin^2\theta/2 \\ 
			\frac{1}{\sqrt{2}}\sin\theta \qquad  \cos\theta  \qquad  -\frac{1}{\sqrt{2}}\sin\theta \\
			\sin^2\theta/2 \qquad \frac{1}{\sqrt{2}}\sin\theta \qquad \cos^2\theta/2
		\end{pmatrix}\,.
	\end{aligned}
\end{eqnarray}

The explicit calculation of the AM transfer relation during NBW scattering involves coupling the phase factors that depend on the azimuth angle $\varphi$ and encode SAM and OAM contributions of the corresponding particles, in Eqs.~\eqref{eq_a}, \eqref{harmonic_expansion}, and \eqref{SAM_basis_expansion}. For the kinematic scenario of $\theta_{\bm{p}_{e^-}}=0$ [illustrated in Fig.~\ref{fig_1} (b)], the scattering process exhibits rotational symmetry in the transverse plane, and the azimuthal angle dependent phase factors turn into a delta function that represents AM conservation. In the following, we obtain the AM-resolved NBW scattering amplitude and probability distributions using the previously introduced definitions.\\ 

\subsubsection{Probability for the detection of vortex positron}

The $S$-matrix element for the generation of vortex positron from the vortex $\gamma$ photon (\rm $\gamma_{\text{vortex}}$ + $n\,\omega_L$ $\rightarrow$ $e^+_{\text{vortex}}$ + $e^-_{\text{plane wave}}$) can be expressed as
\begin{equation}
	\begin{aligned}
		S_{fi}^\text{2vortex}=&\int\frac{d^2\bm{k}_{\gamma,\perp}}{(2\pi)^2}\frac{d^2 \bm{p}_{{e^+},\perp}}{(2\pi)^2}		a_{j_\gamma}(\bm{k}_{\gamma,\perp})\,a^\ast_{j_{e^+}}(\bm{p}_{{e^+},\perp})\,S_{fi}^\text{plane}\,\\
		=&\frac{ie(2\pi)}{\omega_L}\delta({k}_{\gamma,z}+s{k}_{L,z}-{p}_{{e^-},z}-{p}_{{e^+},z})\delta(p_{\perp0}-k_{\perp0})\,\\
		& \times\sum_{n}\,i^{j_{\gamma}+j_{{e^+}}}\,\delta_{j_{\gamma}+n,\,j_{{e^+}}+s_{e^-}}\mathcal{M}_{n}(s)\,,
	\end{aligned}\label{eq_vorpair}
\end{equation}
where $a_{j_\gamma}(\bm{k}_{\gamma,\perp})$ and $a^\ast_{j_{e^+}}(\bm{p}_{{e^+},\perp})$ denote vortex amplitudes of the vortex $\gamma$ photon (with TAM $j_\gamma$) and the vortex positron (with TAM $j_{e^+}$), respectively. The quantity $s=\frac{k_\gamma p_{e^+}}{k_L (k_\gamma-p_{e^+})}$ is the continues laser photon absorption number obtained from the energy-momentum conservation relation. In Eq.~\eqref{eq_vorpair}, we project the final states onto a plane-wave electron and a vortex positron, which amounts to postselecting a plane-wave electron and detecting the vortex positron. Upon postselecting the electron alone would result in the generation of vortex positrons in a superposition of different OAM modes, as shown in Eq.~\eqref{pos_NBW_ev} in Appendix \ref{Appendix_B}. The spin and helicity dependent probability for detecting vortex positrons with energy $\varepsilon_{e^+}$ is given by
\begin{equation}
		\frac{dW^{\text{2vortex}}}{d\varepsilon_{e^+}}=\frac{\alpha}{4\varepsilon_\gamma}\frac{\varepsilon_{e^+}p_{e^+,z}}{(k_L\,p_{e^-})(k_L\,p_{e^+})}\sum_n\delta_{j_{\gamma}+n,j_{e^+}+s_{e^-}}\vert\,\mathcal{M}_{n}(s)\vert^2\,,
		\label{eq_NBW_vortex}
\end{equation}
where we assume the special condition for the electron's polar angle $\theta_{\bm{p}_{e^-}}=0$. The amplitude $\mathcal{M}_n$ is given by
\begin{equation}
	\begin{aligned}
		&\mathcal{M}_n=\\
		&d^{1/2}_{s_{e^+},s_{e^-}}(\theta_{1})d^{1}_{2s_{e^-},\Lambda}(\theta_{k})(-\sqrt{2})f^{(2)}\mathscr{C}^{(n)}_{0}(s)-d^{1/2}_{-s_{e^+},s_{e^-}}(\theta_{1})
		d^{1}_{0,\Lambda}(\theta_{k})\\&\times[f^{(2)}\mathscr{C}^{(n)}_{0}(s)+2\alpha_{e^-}\alpha_{e^+}(-2s_{e^+}f^{(1)}+f^{(2)})\mathscr{C}^{(n)}_{2}(s)]+\\
		&\mathscr{C}^{(n)}_{2s_{e^+}}(s)\{d^{1/2}_{s_{e^+},s_{e^-}}(\theta_{1})d^{1}_{0,\Lambda}(\theta_{k})[\alpha_{e^-}(f^{(1)}+2s_{e^+}f^{(2)})+\\
		&\alpha_{e^+}(f^{(1)}-2s_{e^-}f^{(2)})]+\\
		&d^{1/2}_{-s_{e^+},s_{e^-}}(\theta_{1})d^{1}_{-2s_{e^+},\Lambda}(\theta_{k})\sqrt{2}\alpha_{e^+}(f^{(1)}-2s_{e^+}f^{(2)})\}+\\
		&\mathscr{C}^{(n)}_{-2s_{e^+}}(s)d^{1/2}_{-s_{e^+},s_{e^-}}(\theta_{1})d^{1}_{2s_{e^+},\Lambda}(\theta_{k})\sqrt{2}\alpha_{e^-}(f^{(1)}-2s_{e^+}f^{(2)}).
	\end{aligned} 
	\label{eq_NBW_vortex_M}
\end{equation}
For better readability, we use $\theta_{1}$ and $\theta_k$ as short notations for $\theta_{\bm{p}_{e^+}}$ and $\theta_{\bm{k}_\gamma}$, respectively. The following definitions are used: 
\begin{eqnarray}
	\begin{aligned}
		f^{(1)}&=2s_{e^-}\sqrt{\varepsilon_{e^-}-m_{e}}\sqrt{\varepsilon_{e^+}+m_{e}}-2s_{e^+}\sqrt{\varepsilon_{e^-}+m_{e}}\sqrt{\varepsilon_{e^+}-m_{e}}\,,\\
		f^{(2)}&=\sqrt{\varepsilon_{e^-}+m_{e}}\sqrt{\varepsilon_{e^+}+m}+2s_{e^-}(-2s_{e^+})\sqrt{\varepsilon_{e^-}-m_{e}}\sqrt{\varepsilon_{e^+}-m_{e}}\,,\nonumber
	\end{aligned}
\end{eqnarray}
and the dynamic integrals are
\begin{eqnarray}
	\begin{aligned}
		\mathscr{C}_{0}^{(n)}(s)&=\int d\phi\,e^{i(s-n)\phi-i(\beta_{e^-}+\beta_{e^+})\int^{\phi}d \phi^{\prime}g^{2}(\phi^{\prime})}J_{n}(\alpha_{p_{e^+}}g(\phi))\,,\nonumber\\
		\mathscr{C}^{(n)}_{2}(s)&=\int d\phi\,e^{i(s-n)\phi-i(\beta_{e^-}+\beta_{e^+})\int^{\phi}d \phi^{\prime}g^{2}(\phi^{\prime})}g^{2}(\phi)J_{n}(\alpha_{p_{e^+}}g(\phi))\,,\nonumber\\
		\mathscr{C}^{(n)}_{\sigma=\pm}(s)&=\int d\phi\,e^{i(s-n)\phi-i(\beta_{e^-}+\beta_{e^+})\int^{\phi}d \phi^{\prime}g^{2}(\phi^{\prime})}g(\phi)J_{n-\sigma}(\alpha_{p_{e^+}}g(\phi))\,,
	\end{aligned}\label{dyn_int_har}
\end{eqnarray}
with the kinematic factors defined as
\begin{eqnarray}
	\begin{aligned}
		\alpha_{p_{e^+}}=\frac{m_e\,p_{{e^+}\perp}}{k_Lp_{e^+}}a_0,\,\,\,\alpha_{{{e^-}/{e^+}}}=\frac{m_e\,\omega_L}{k_Lp_{{e^-}/{e^+}}}a_0,\,\,\,\beta_{{e^-}/{e^+}}=\frac{m_e^2a_0^2}{2k_Lp_{{e^-}/{e^+}}}\,.\nonumber
	\end{aligned}
\end{eqnarray}

In Eq.~\eqref{eq_NBW_vortex}, the interference term $\sum_{n\neq n^\prime}\mathcal{M}_{n}\mathcal{M}_{n^\prime}^\ast$ is eliminated by the AM conserving $\delta$ function, which enables one to express the TAM of the created vortex positron as $j_{e^+}=j_\gamma+n-s_{e^-}$. Thus, the vortex positron receives AM contributions from the $\gamma$ photon ($j_\gamma$), the multiple laser photons ($n$), and the SAM of the electron ($s_{e^-}$).  Here, laser photons impart extra AM into the scattering process, enhancing the contribution from the vortex $\gamma$ photon $j_\gamma$ by $n$. Therefore, higher particle yield, plane-wave postselection criteria, and AM enhancement via multiphoton absorption, facilitated by the intense laser, distinguish our results from linear scattering processes \cite{Bu:2021ebc,Lei:2021eqe,Lei:2023wui}. \\

\subsection{Numerical results}

\begin{figure}[H]
	\setlength{\abovecaptionskip}{-0.0cm}
	\includegraphics[width=1\linewidth]{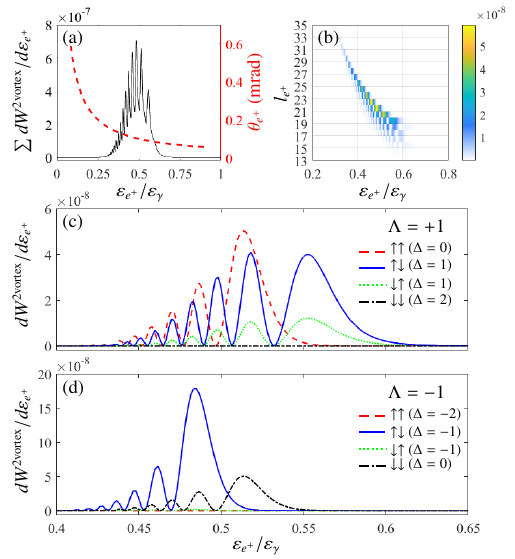}
	\begin{picture}(300,0)		
	\end{picture}
	\caption{(a) Solid black line: the probability distribution of the vortex positron generation with respect to the energy ratio $\varepsilon_{e^+}/\varepsilon_\gamma$. The probability is summed for NBW harmonics and particle spins. Dashed red line: the vortex positron's cone opening angle $\theta_{\bm{p}_{e^+}}$ vs the energy ratio $\varepsilon_{e^+}/\varepsilon_\gamma$. (b) The probability distribution of the vortex positron generation with respect to OAM $l_{e^+}$ and the energy ratio $\varepsilon_{e^+}/\varepsilon_\gamma$ for helicity $\Lambda=1$ and spins $s_{e^-}=s_{e^+}=1/2$. (c),(d): Probability distributions of the vortex positron generation with respect to the energy ratio $\varepsilon_{e^+}/\varepsilon_\gamma$ for the fixed $l_{e^+}=20$ for the $\gamma$ photon helicities $\Lambda=+1$ and $\Lambda=-1$, respectively. The arrows on the left (right) represent the electron (positron) spin states, with $\uparrow$ and $\downarrow$ denoting $s_{e^-/e^+}=\frac{1}{2}$ and $s_{e^-/e^+}=-\frac{1}{2}$, respectively. The spin contributions from scattering particles are defined as $\Delta=\Lambda-s_{e^-}-s_{e^+}$. }
	\label{fig_3}
\end{figure}

The energy spectrum of the vortex positron is determined numerically by utilizing Eq.~\eqref{eq_NBW_vortex}. We display the results for a vortex $\gamma$ photon with  cone opening angle of $\theta_{\bm{k}_\gamma}=0.05$ mrad while employing the same parameters as in Fig.~\ref{fig_CP}. After examining a broad spectrum of harmonics spanning from 1 to 80, we found that the most significant contributions come from those within the range $10 \leq n \leq 40$. The cumulative effect of all other harmonics outside this range is at least 4 orders of magnitude smaller. To better demonstrate the OAM distribution, the range of the plots has been made smaller and now includes only the interval $12 \leq n \leq 35$. We find that the dominant contribution comes from harmonics in the range $10\leq n\leq35$. As shown in Fig.~\ref{fig_3} (a), the vortex positron is produced with a broad energy spectrum due to intense laser pulses. Given the relation $\vert\bm{p}_{{e^+},\perp}\vert=\vert\bm{k}_{\gamma,\perp}\vert$, the energy of the vortex positron $\varepsilon_{e^+}$ and its cone opening angle $\theta_{\bm{p}_{e^+}}=\arcsin{\vert \bm{p}_{{e^+}\perp} \vert}/{\vert \bm{p}_{e^+}\vert}$ satisfy $\varepsilon_{e^+}\approx~\vert\bm{k}_{\gamma,\perp}\vert/\theta_{\bm{p}_{e^+}}$ [the dashed red line in Fig.~\ref{fig_3} (a)]. Moreover, considering the kinematic condition $\theta_{\bm{p}_{e^-}}=0$, specifying the energy $\varepsilon_{e^+}$ of the positron uniquely determines its polar angle $\theta_{\bm{p}_{e^+}}$, thereby defining a conical shape in momentum space. The remaining aspect to be characterized is its AM property.\\

Both the AM of the absorbed laser photons and the spin contributions from the scattering particles play significant roles in determining the AM of the vortex positron.  
In our collision scenario, with particle energies in the multi-GeV range and transverse momenta below a few MeV, the paraxial approximation holds, rendering the intrinsic spin-orbit interaction associated with the vortex particle negligible. Consequently,the TAM of the vortex particle can be decomposed into the SAM and OAM \cite{Leader:2013jra,Leader:2015vwa}. The OAM of the vortex $\gamma$ photon is given by $l_\gamma=j_\gamma-\Lambda$, while that of the vortex positron is $l_{e^+}=j_{e^+}-s_{e^+}$, which results in the relation $l_{e^+}=l_\gamma+n+\Delta$, with $\Delta=\Lambda-s_{e^-}-s_{e^+}$. According to Eq.~\eqref{eq_NBW_vortex}, the generation probability of vortex positrons involves incoherent summations over various combinations of spin and helicity, as well as different harmonic orders. For a given OAM $l_\gamma=1$ and specific helicity ($\Lambda=1$) and spin ($s_{e^-}=s_{e^+}=1/2$) configurations, the positron OAM $l_{e^+}=1+n$ is uniquely determined for each harmonic $n$ [Fig.~\ref{fig_3} (b)]. Here, the $n$th harmonic contributes an AM number of $n$ to the positron. Various spin and helicity configurations of the NBW process contribute to the same OAM value of the vortex positron. In this scenario, $\Delta$ encapsulates spin contributions from scattering particles and assumes integer values within the range $-2\leq \Delta \leq 2$. In Figs.~\ref{fig_3} (c) and 5(d), the pair creation rates for eight distinct spin configurations are depicted for $l_{e^+}=20$. Among these, two configurations are significantly suppressed, and they correspond to $\Delta=\pm 2$. Consequently, the contributing values of $\Delta$ are limited to $\{-1, 0, 1\}$. Given that a substantial number of laser photons ($n\gg\Delta$) participate in the pair creation process, the OAM of the vortex positron arises from a transfer of OAM from the initial $\gamma$ photon and the AM of multiple laser photons, leading to $l_{e^+}\approx l_\gamma +n$.\\

\begin{figure}[H]
	\setlength{\abovecaptionskip}{-0.0cm}
	\includegraphics[width=0.975\linewidth]{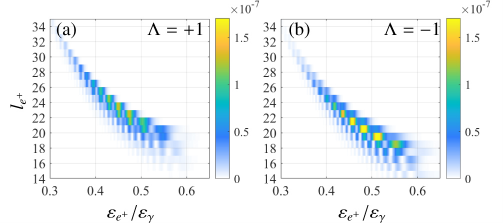}
	\begin{picture}(300,0)	
	\end{picture}
	\caption{The probability distribution of the vortex positron generation with respect to the positron's OAM and the energy ratio $\varepsilon_{e^+}/\varepsilon_\gamma$.The $\gamma$ photon carries OAM $l_\gamma=1$ and its helicity takes the values (a)$\Lambda=+1$ and (b) $\Lambda=-1$. The same set of $\gamma$ photon and laser pulse parameters is used as in Fig.~\ref{fig_3}. } 
	\label{fig_4}
\end{figure}  

The OAM- and energy-resolved probability distributions for the generated vortex positron are presented in Fig.~\ref{fig_4}. Given the specified parameters $\omega_L=1.55$ eV and $\omega=20$ GeV, it is estimated that the number of absorbed laser photons exceeds $n\gtrsim17$. As a result, even when the initial $\gamma$ photon has $l_\gamma=1$, the resulting positron attains a substantial OAM (Fig.~\ref{fig_4}). Upon comparing the results in Figs.~\ref{fig_4} (a) and 6(b), it becomes apparent that the vortex $\gamma$ photon with negative helicity ($\Lambda=-1$) is more favorable, exhibiting a relatively higher probability than its positive helicity counterpart. Notably, the OAM of the $\gamma$ photon and AM of the multiple laser photons are transferred solely to the positron, leaving the electron as a plane-wave particle.  This scenario differs from NBW pair creation processes involving extrinsic OAM, where both the electron and the positron acquire the same amount of AM \cite{Chen:2018tkb,Zhu:2018und,Zhao2022,Zhang:2024ofv}.  

\section{discussion}
\subsection{Collision of the vortex $\gamma$ photon with the intense laser}
Considering the experimental feasibility of our proposal, the on-axis scattering geometry presents significant challenges both from an experimental standpoint and in terms of measuring the final particles. For scenarios involving on-axis or minimally angled scattering of the final electron, the magnitudes of the transverse momenta of involved particles adhere to relations such that \(|\bm{k}_{\gamma,\perp}| \approx |\bm{p}_{{e^+},\perp}| \lesssim 1 \, \text{MeV} \gg |\bm{p}_{{e^-},\perp}|\). Assuming typical particle energies in the GeV range, the implied polar angles are \(\theta_{\gamma} \approx \theta_{e^+} \lesssim 1\) mrad and \(\theta_{e^-} \ll 1\)mrad. Consequently, experimentally discerning the produced vortex leptons becomes exceedingly challenging due to these minute angular separations. Concerning the assumption of an incoming vortex $\gamma$ photon, there are theoretical proposals showing the possibility of producing vortex $\gamma$ photons  via Compton scattering processes \cite{jentschura2011generation,taira2017gamma,bogdanov2019semiclassical,Karlovets:2021gcm,Ababekri:2022mob,Guo:2023uyu}. The pair creation setup is considered for the multiphoton regime ($a_0\gtrsim1$) \cite{Burke:1997ew,Hu:2010ye}, with $\gamma$ photon energies typically ranging around $\varepsilon_\gamma\sim$ GeV. These parameters align with those used in the E144 experiment \cite{Burke:1997ew} and are also in line with those planned for upcoming experimental setups \cite{Macleod:2021xeq,Salgado:2021fgt,Abramowicz:2021zja}. \\

For GeV photons, the energy resolution of the operating Compton-\(\gamma\) sources has already reached \(\frac{\Delta \varepsilon_\gamma}{\varepsilon_\gamma}\lesssim1.25\%\) \cite{Schaerf:2005um,Weller2009}. In linear Compton sources, $\gamma$ photons are predominantly generated along the scattering axis with the energy $\varepsilon_\gamma\approx4\gamma_e^2\omega_L$ and the angular spread $\Delta\theta\lesssim\frac{1}{\gamma_e^2}$, where $\gamma_e$ is the electron Lorentz factor \cite{jentschura2011generation,Karlovets:2021gcm}. Moreover, nonlinear Compton sources with intense laser pulses can offer higher brilliance exceeding $10^{21}$ photons s$^{-1}$ mm$^{-2}$ mrad$^{-2} 0.1$\% BW at the cost of energy and divergence broadening due to multiphoton absorption and quantum stochasticity effects \cite{Tang:2020xlj,Fedotov:2022ely}, which may be narrowed by chirping \cite{Seipt:2019yds,app13020752}. In theory, pair creation is also feasible using $\gamma$ photons of lower energies. To achieve this, one could either increase the laser intensity (thus increasing the value of $a_0$) or employ lasers with higher frequencies, such as free-electron laser/x-ray free-electron laser pulses, to achieve substantial quantum nonlinearity parameter $\chi=a_0\frac{2\omega_L\varepsilon_\gamma}{m_e^2}$ \cite{Fedotov:2022ely}.  However, it is crucial to note that employing ultraintense lasers ($a_0 \gg 1$) may initiate multiple radiation emissions and pair creations. These complex events must be carefully considered in the context of vortex particle scattering and could be the subject of separate, focused investigations. \\

Concerning the results for the vortex positron's energy and OAM distributions given in Figs.~\ref{fig_3} and \ref{fig_4}, we consider the NBW scattering of a single \(\gamma\) photon with a laser pulse. We present numerical results for a fixed set of parameters: a vortex \(\gamma\) photon energy \(\varepsilon_\gamma = 20\) GeV and a cone opening angle of \(\theta_{\bm{k}_\gamma} = 0.05\) mrad (corresponding to \(k_{\gamma,\perp} = 1\) MeV). A spread in \(\gamma\) photon energy would alter the numerical results by producing distinct energy and OAM distributions for each set of \(\varepsilon_\gamma\) and \(\theta_{\bm{k}_\gamma}\). However, the mechanism of vortex positron generation allows for variations in the energy spread. The key condition for vortex positron generation with maximum yield when postselecting an electron along the \(z\) axis is \(\theta_{\text{pair}} = \theta_{\bm{k}_\gamma}\). Here, the typical pair creation angle is estimated as \(\theta_{\text{pair}} \sim \frac{m_e a_0}{\varepsilon_\gamma}\), and the cone opening angle is given as \(\theta_{\bm{k}_\gamma} = \frac{k_{\gamma,\perp}}{\varepsilon_\gamma}\). Thus, the condition translates into the relation \(m_e a_0 \sim k_{\gamma,\perp}\), which imposes no specific restriction on the \(\gamma\) photon energy. However, a significant decrease in the \(\gamma\) photon energy would decrease the total pair creation probability while potentially increasing the number of absorbed laser photons. \\

\subsection{Detection of the cone angle and the superposition states of the vortex $\gamma$ photon}

Beyond their intrinsic OAM,  vortex particles exhibit distinctive transverse coherence that manifests in the form of a cone opening angle. This unique feature is anticipated to introduce novel kinematic aspects to scattering phenomena. In the context of our study, the cone opening angle $\theta_{\bm{k}_\gamma}$ of the vortex $\gamma$ photon plays a pivotal role in optimizing the generation of vortex positrons, as exemplified in Fig.~\ref{fig_02}.   We demonstrate the feasibility of detecting this angle by analyzing the created pairs' angular distribution during the NBW process. The pair creation probability in terms of the electron's energy $\varepsilon_{e^-}$ and solid angle $\Omega_{e^-}$ is
\begin{equation}
	\frac{d^{2}W^{\text{1vortex}}}{d\varepsilon_{e^-}d\Omega_{e^-}}=\int\frac{d\varphi_{\bm{k}_\gamma}}{2\pi}F(\varphi_{\bm{k}_\gamma})\frac{d^2W^{\text{plane}}(\theta_{\bm{k}_\gamma},\varphi_{\bm{k}_\gamma})}{d\varepsilon_{e^-}d\Omega_{e^-}}\,.
	\label{eq_vortex_pair_rate_sup}
\end{equation}
Here, $F(\varphi_{{\bm{k}_\gamma}})$ is related to the vortex state of the incoming $\gamma$ photon such that $F(\varphi_{{\bm{k}_\gamma}})=1$ corresponds to the $\gamma$ photon in the vortex eigenstate given by Eq.~\eqref{eq_vortex_pair_rate}, and $F(\varphi_{{\bm{k}_\gamma}})=1+\cos[\Delta j_\gamma(\varphi_{{\bm{k}_\gamma}}-\frac{\pi}{2})+\delta]$ corresponds to the $\gamma$ photon in the vortex superposition state ($\Delta j_\gamma=j_{\gamma,1}-j_{\gamma,2}$). The vortex amplitude of the superposition state is given as
\begin{equation}
	\tilde{a}_{j_{\gamma,1},j_{\gamma,2}}(\bm{k}_{\gamma,\perp})=\frac{1}{\sqrt{2}}[a_{j_{\gamma,1}}(\bm{k}_{\gamma,\perp})+e^{i\delta}a_{j_{\gamma,2}}(\bm{k}_{\gamma,\perp})].
	\label{eq_superposition}
\end{equation}
In our numerical calculations, we have set the relative phase factor to $\delta = 0$. This choice does not affect the conclusions drawn in our study, as varying $\delta$ merely induces a rotation of the angular distribution around the center of the $\theta_{e^-,x}\theta_{e^-,y}$ plane without altering its overall shape.  \\

\begin{figure}[H]
	\setlength{\abovecaptionskip}{-0.0cm}
	\includegraphics[width=0.95\linewidth]{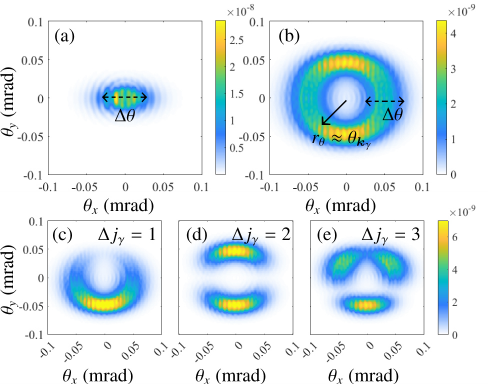}
	\begin{picture}(300,0)	
	\end{picture}
	\caption{The probability distribution $dW^\text{1vortex}({\theta_{\bm{k}_\gamma}})/d\Omega_{e^-}$ of the electron with respect to its propagation angle in the $\theta_{{e^-,x}}\theta_{{e^-,y}}$ plane during NBW scattering of the $\gamma$ photon in LP lasers. The $\gamma$ photon is in (a) a plane-wave state and (b) a vortex state. $\gamma$ photons are in the vortex superposition states with $\Delta j_\gamma=1$,2,and 3 in (c), (d), and (e), respectively. } 
	\label{fig_5}
\end{figure} 
 
The numerical results are obtained by using Eq.~\eqref{eq_vortex_pair_rate}  
for a linearly polarized (LP) laser pulse $A^\mu(\phi)=a\,g(\phi)\left\{0,\cos\phi,0,0 \right\}$ with intensity $a_0=1$ for the same vortex $\gamma$ photon parameter as in Fig.~\ref{fig_3}. 
When the incoming $\gamma$ photon is in the plane-wave state ($\theta_{\bm{k}_\gamma}=0$), the created electron is concentrated at the center and exhibits a diameter $\Delta\theta$ that corresponds to the FWHM of the electron's angular distribution in the $\theta_{{e^-,x}}\theta_{{e^-,y}}$ plane [Fig.~\ref{fig_5} (a)]. The vortex state of the $\gamma$ photon causes the angular distribution of the electron to spread along a circle with radius $r_{\theta}$ and thickness $\Delta\theta$ [Fig.~\ref{fig_5} (b)]. $r_\theta$ represents the radial distance from the center to the peak position of the probability distribution ring in the $\theta_{{e^-,x}}\theta_{{e^-,y}}$ plane. We observe that the radius of the electron angle distribution $r_{\theta}$ is closely related to the polar angle of the vortex $\gamma$ photon $\theta_{\bm{k}_\gamma}\approx\,r_{\theta}$; thus, one can discern the polar angle $\theta_{\bm{k}_\gamma}$ by measuring $r_{\theta}$.  However, for the polar angle to be clearly resolved, the process should satisfy $r_{\theta}\gtrsim\Delta\theta$; hence, one obtains the condition $\theta_{\bm{k}_\gamma}\gtrsim\Delta\theta$. When one estimates $\Delta\theta\sim a_0 m_e/\varepsilon_\gamma$ \cite{Baier:1998vh}, the condition for effective measurement poses an upper limit on the laser intensity of $a_0\lesssim \vert\bm{k}_{\gamma,\perp}\vert/m_e$.\\

Furthermore, when the vortex $\gamma$ photon exists in a superposition state, e.g., with two distinct TAM values $j_{\gamma,1}$ and $j_{\gamma,2}$, as defined in Eq.~\eqref{eq_superposition}---a scenario possible in the generation of vortex $\gamma$ photons via nonlinear Compton scattering in pulsed lasers \cite{Ababekri:2022mob,Guo:2023uyu}---discrete patterns arise that are associated with the difference $\Delta j_\gamma = j_{\gamma,1} - j_{\gamma,2}$, which appears in the function $F(\varphi_{\bm{k}_\gamma})$ in Eq.~\eqref{eq_vortex_pair_rate_sup}.  In this case, the electron angle distribution can still reveal the polar angle $\theta_{\bm{k}_\gamma}$ as pairs are created along the circle with radius $r_{\theta}\approx\theta_{\bm{k}_\gamma}$ [Figs.~\ref{fig_5} (c) and 7(d)].\\

\begin{figure}[H]
	\setlength{\abovecaptionskip}{-0.0cm}
	\includegraphics[width=0.95\linewidth]{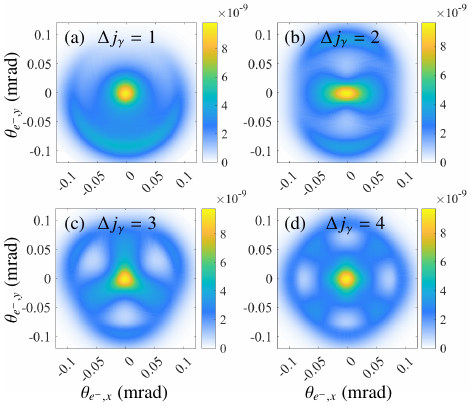}
	\begin{picture}(300,0)	
	\end{picture}
	\caption{The probability distribution ${dW^{\text{1vortex}}}/{d\Omega_{e^-}}$ of the electron in $\theta_{e^-,x}\theta_{e^-,y}$ plane corresponding to NBW scattering in CP lasers with a $\gamma$ photon in the vortex superposition states with different $\Delta j_\gamma$ values.} 
	\label{fig_sup_CP}
\end{figure}

The probability distribution ${dW^{\text{1vortex}}}/{d\Omega_{e^-}}$ of the created electron in the $\theta_{e^-,x}\theta_{e^-,y}$ plane is shown in Fig.~\ref{fig_sup_CP} for vortex superposition states of the $\gamma$ photon in the CP laser case.  Discrete patterns related to the value of $\Delta j_\gamma$ arise in Fig.~\ref{fig_sup_CP} as in the corresponding cases for a LP laser pulse in Figs.~\ref{fig_5} (c)and 7(d).
Comparing Figs.~\ref{fig_CP} and \ref{fig_sup_CP} for the CP laser to the corresponding results in Fig.~\ref{fig_5} for the LP laser, we conclude that the LP laser case yields clearer signatures and is  more suitable for detecting the cone opening angle $\theta_{\bm{k}_\gamma}$ of the vortex $\gamma$ photon.\\

\section{conclusion}

In conclusion, we put forward a novel method of generating ultrarelativistic vortex positrons and electrons with large OAM through NBW scattering of vortex $\gamma$ photon in CP lasers. We developed a complete AM-resolved quantum scattering theory and reveal the OAM-transfer mechanism of an NBW process. Our theoretical framework can be further developed to investigate vortex particle generation and detection in other SF-QED processes, and our findings also pave the way for the application of ultrarelativistic vortex leptons in high-energy particle physics, nuclear physics, astrophysics, etc.

\begin{flushright}
\end{flushright}

{\it ACKNOWLEDGMENTS} The work is supported by the National Natural Science Foundation of China (Grants No.U2267204, No.12105217, and No.12147176), the Foundation of Science and Technology on Plasma Physics Laboratory (Grant No.JCKYS2021212008), the Natural Science Basic Research Program of Shaanxi (Grants No.2023-JC-QN-0091 and No.2024JC-YBQN-0042), and the Shaanxi Fundamental Science Research Project for Mathematics and Physics (Grants No. 22JSY014 and No.22JSQ019).

\appendix

\section{POSTSELECTING A BESSEL VORTEX ELECTRON}\label{Appendix_A}

We begin by deriving the probability rate for the case where both the electron and positron are vortex particles. Subsequently, we calculate the result for the scenario where the electron is postselected in the zero-Bessel mode. The $S$matrix of the process (\rm $\gamma_{\text{vortex}}$ + $n\,\omega_L$ $\rightarrow$ $e^+_{\text{vortex}}$ + $e^-_{\text{vortex}}$) is written as
\begin{equation}
	\begin{aligned}
		S_{fi}^\text{3vortex}=&\int\frac{d^2\bm{k}_{\gamma\perp}}{(2\pi)^2}\frac{d^2\bm{p}_{1\perp}}{(2\pi)^2}\frac{d^2 \bm{p}_{2\perp}}{(2\pi)^2}\\&\times~a_{j_\gamma}(\bm{k}_{\gamma,\perp})\,a^\ast_{j_1}(\bm{p}_{1,\perp})\,a^\ast_{j_2}(\bm{p}_{2,\perp})\,S_{fi}^\text{pl}\,,
	\end{aligned}\label{vortex_vortex}
\end{equation}
where the initial $\gamma$ photon, final electron, and positron are all vortex particles. As both of the final state particles in Eq.~\eqref{vortex_vortex} are projected onto the vortex modes, this corresponds to the assumption of postselecting the final electron (or positron) via a twisted detector as in \cite{Bu:2021ebc,Lei:2021eqe,Lei:2023wui}. The notations for the positron and electron are used as follows: $p_1 = p_{e^+}$, $p_2 = p_{e^-}$ for their momenta, and $j_1 = j_{e^+}$, $j_2 = j_{e^-}$ for their TAM.  The pair creation rate reads
\begin{equation}
	\begin{aligned}
		\frac{d^3W}{dp_{1\perp}dp_{1z}dp_{2\perp}}&=\frac{\pi e^2}{8}\frac{k_{\perp0}~p_{1\perp0}p_{2\perp0}}{(k_L\,p_1)(k_L\,p_2)}\vert \sum_{n_1,n_2}\delta_{j_{\gamma}+\Delta n,j_1+j_2}\\
		&\times\sum_{\sigma\dots}~\frac{\cos(\Delta{m}_1\delta_1-\Delta{m}_2\delta_2)}{p_{1\perp}p_{2\perp}\sin(\delta_1+\delta_2)}\mathcal{M}_{n_1,n_2}\vert^2,\end{aligned}
	\label{eq_NBW_vortex3}
\end{equation}
where the definitions of new symbols are $\Delta n=n_1-n_2$, $\Delta m_{1/2}=j_{1/2}-\tilde{m}_{1/2}$, $\tilde{m}_1=n_1-\sigma+\sigma_{1}-s_1$, $\tilde{m}_2=-n_2+\sigma_{2}-s_2$, $\delta_1=\arccos\frac{k_{\perp0}^2+p_{1\perp0}^2-p_{2\perp0}^2}{2k_{\perp0}p_{1\perp0}}$, and $\delta_2=\arccos\frac{k_{\perp0}^2-p_{1\perp0}^2+p_{2\perp0}^2}{2k_{\perp0}p_{2\perp0}}$ \cite{Ivanov:2011kk}. In the second line of Eq.~\eqref{eq_NBW_vortex3}, the summations over dummy indices $\{\sigma, \sigma_{1}, \sigma_{2}, \sigma_{\gamma}\}$ are performed.  The harmonics are defined as follows:
\begin{eqnarray}
	\begin{aligned}
		\mathcal{M}_{n_1,n_2}=&[\delta_{\sigma,0}(\delta_{\sigma_{1},\sigma_{2}}\mathcal{G}_{0}^{\uparrow\uparrow}+\delta_{\sigma_{1},-\sigma_{2}}\mathcal{G}_{0}^{\uparrow\downarrow})-(\delta_{\sigma_{1},\sigma_{2}}\mathcal{G}_{\pm}^{\uparrow\uparrow} \\ &+\delta_{\sigma_{1},-\sigma_{2}}\mathcal{G}_{\pm}^{\uparrow\downarrow})]d^{1/2}_{\sigma_1,s_1}(\theta_{\bm{p}_{1}})d^{1/2}_{\sigma_{2},s_2}(\theta_{\bm{p}_{2}})d^{1}_{\sigma_{\gamma},\Lambda}(\theta_{\bm{k}}),
	\end{aligned}
	\label{eq_NBW_vortex3_M}
\end{eqnarray}
with the following notations: 
\begin{eqnarray}
	\begin{aligned}
		\mathcal{G}_{0}^{\uparrow\uparrow}=&\delta_{\sigma_{\Lambda},2\sigma_{1}}(-\sqrt{2})f^{(2)}\,\mathscr{C}_{0}^{(n_1, n_2)}(s)\,,\\
		\mathcal{G}_{0}^{\uparrow\downarrow}=&\delta_{\sigma_{\Lambda},0}[f^{(2)}\,\mathscr{C}_{0}^{(n_1, n_2)}(s)\\&+2\alpha_{1}\alpha_{2}(2\sigma_{1} f^{(1)}+f^{(2)})\,\mathscr{C}_{2}^{(n_1, n_2)}(s)]\,,\\		
		\mathcal{G}_{\pm}^{\uparrow\uparrow}=&\delta_{\sigma,2\sigma_{1}}\delta_{\sigma_{\Lambda},0}[\alpha_{1}(f^{(1)}+2\sigma_1 f^{(2)})\,\\
		&+\alpha_{2}(f^{(1)}-2\sigma_1 f^{(2)})]\,\mathscr{C}_{\sigma}^{(n_1, n_2)}(s)\,,\\
		\mathcal{G}_{\pm}^{\uparrow\downarrow}=&[\delta_{\sigma,2\sigma_{1}}\delta_{\sigma_{\Lambda},-2\sigma_{1}}\sqrt{2}\alpha_{1}(f^{(1)}+2\sigma_1 f^{(2)})\,\\
		&+\delta_{\sigma,-2\sigma_{1}}\delta_{\sigma_{\Lambda},2\sigma_{1}}\sqrt{2}\alpha_{2}(f^{(1)}+2\sigma_1 f^{(2)})]\,\mathscr{C}_{\sigma}^{(n_1, n_2)}(s)
		\,.
	\end{aligned}
\end{eqnarray}

The dynamic integrals are written as follows:
\begin{eqnarray}
	\begin{aligned}
		\mathscr{C}_{0}^{(n_1,n_2)}(s)&=\int d\phi\,e^{i(s-\Delta n)\phi-i(\beta_1+\beta_2)\int^{\phi}d \phi^{\prime}g^{2}(\phi^{\prime})}\\ &\times J_{n_1}(\alpha_{p_1}~g(\phi))J_{n_2}(\alpha_{p_2}~g(\phi))\,,\\
		\mathscr{C}^{(n_1,n_2)}_{2}(s)&=\int d\phi\,e^{i(s-\Delta n)\phi-i(\beta_1+\beta_2)\int^{\phi}d \phi^{\prime}g^{2}(\phi^{\prime})}\\ &\times J_{n_1}(\alpha_{p_1}~g(\phi))J_{n_2}(\alpha_{p_2}~g(\phi))g^2(\phi)\,,\\
		\mathscr{C}^{(n_1,n_2)}_{\sigma=\pm}(s)&=\int d\phi\,e^{i(s-\Delta n)\phi-i(\beta_1+\beta_2)\int^{\phi}d \phi^{\prime}g^{2}(\phi^{\prime})}\\ &\times J_{n_1-\sigma}(\alpha_{p_1}~g(\phi))J_{n_2}(\alpha_{p_2}~g(\phi))g(\phi)\,.
	\end{aligned}
\label{dyn_int_2har}
\end{eqnarray}

In the main text, we assumed the zero scattering angle ($\theta_{2}=0$) for the electron and derived the AM conservation relations in Eqs.~\eqref{eq_vorpair} and \eqref{eq_NBW_vortex}. To incorporate a more realistic scenario with a very small scattering angle $\theta_{2}\rightarrow0$, we consider postselecting an electron with vanishing transverse momentum and an undetermined azimuthal angle, which we model using a cone structure with a very small transverse momentum radius $p_{2\perp0}$ that is akin to a zero-Bessel mode. 
For the case of an electron postselected with a zero-Bessel mode, we equate its TAM to its spin, $j_{2} = s_{2}$, in Eq.~\eqref{eq_NBW_vortex3}. Consequently, we recover the delta function $\delta_{j_\gamma + n,\,j_{1}+s_{2}}$, which is the same as for the postselecting plane-wave electron along the collision axis, as given in Eq.~\eqref{eq_NBW_vortex}. Furthermore, by postselecting electrons that meet the criterion of a small transverse momentum \(\vert \bm{p}_{2\perp}\vert\ll k_{\perp0}\), the resulting vortex positron beam is also produced with a range of indefinite transverse momenta, thereby placing the beam in a mixed state configuration.\\

\section{WAVE FUNCTION OF POSITRONS AFTER POSTSELECTING ELECTRONS ALONG THE SCATTERING AXIS}\label{Appendix_B}

When the final electron is postselected into a zero-Bessel mode, the absolute value of its transverse momentum becomes fixed due to $\delta(|\bm{p}_{2\perp}|-p_{2\perp0})$, as evident from the vortex wave packet amplitude given by Eq.~\eqref{eq_a}. Its azimuthal angle remains undetermined, a situation that is similar to a generalized measurement \cite{Karlovets:2022evc}. This type of generalized measurement is effectively realized when the transverse momentum is exceedingly small $p_{2\perp}\ll k_{\perp}$, making it practically impossible to measure the azimuthal angle $\varphi_{2}$. This scenario aligns naturally with our assumption of electron scattering along the collision axis $\theta_{2}\rightarrow0$. Under these conditions, the wave function of the positron can be derived by computing the evolved state \cite{Berestetskii1982quantum,Karlovets:2022evc}, 
\begin{equation}
	\begin{aligned}
		\vert \psi^{\text{ev}}_{1} \rangle&=\int\frac{d^2\bm{k}_{\gamma\perp}}{(2\pi)^2}\frac{d^2\bm{p}_{2\perp}}{(2\pi)^2}a_{j_\gamma}(\bm{k}_{\gamma,\perp})\,a^\ast_{j_2=0}(\bm{p}_{2,\perp})v_{s_1}S_{fi}^{\text{plane}}\\
		&=\frac{(-ie)}{\omega_L}\delta(k_{\gamma z}+sk_{Lz}-p_{1z}-p_{2z})\sqrt{k_{\gamma\perp}p_{1\perp}}\delta(p_{1\perp}-p_{1\perp0})\\
		&\sum_{n_1,n_2,\sigma,\dots}(-i)^{j_\gamma}e^{-i(j_\gamma-\sigma_\Lambda+\tilde{m}_1+\tilde{m}_2+\sigma^\ast_1)}d^{1/2}_{\sigma^\ast_1,s_1}V_{\sigma^\ast_1}^{s_1}\mathcal{M}_{n_1,n_2}\frac{\cos[\dots]}{\Delta},
	\end{aligned}
	\label{pos_NBW_ev}
\end{equation}  
where the last term on the third line reads $\frac{\cos[(\sigma_\Lambda-j_\gamma-\tilde{m}_2)\delta_1-\tilde{m}_2\delta_2]}{p_{1\perp}p_{2\perp}\sin(\delta_1+\delta_2)/2}$ and the positron spinor is expanded as $v_{s_1}=\sum_{\sigma_1^\ast}e^{-i\sigma_1^\ast\varphi_1}d^{1/2}_{\sigma_1^{\ast},s_1}V_{\sigma_1^\ast}^{s_1}$, as in Eq.~\eqref{SAM_basis_expansion}.  After performing the summation and taking the Kronecker deltas in Eq.~\eqref{eq_NBW_vortex3_M} into account, we are left with the summation over $n_1, n_2,$ and $\sigma_1^\ast$, accompanied by the vortex phase term $e^{-i(j_\gamma+\Delta n-s_1+\sigma_1^\ast)\varphi_1}$. Therefore, Eq.~\eqref{pos_NBW_ev} represents a superposition of vortex modes with various $\Delta n$ values, with each mode carrying TAM $j_1 = j_\gamma + \Delta n - s_2$. \\

\vskip 1.0cm

\bibliography{NBW_blib}

\end{document}